\documentclass[12pt,preprint]{aastex}
\newcommand{\beq}{\begin{equation}}
\newcommand{\beqa}{\begin{eqnarray}}
\newcommand{\eeq}{\end{equation}}
\newcommand{\eeqa}{\end{eqnarray}}

\newcommand{\tot}{{\rm tot}}
\newcommand{\obs}{{\rm obs}}
\newcommand{\sub}{{\rm sub}}
\newcommand{\dep}{{\rm dep}}

\def\j{{\scriptscriptstyle (j)}}
\def\tot{{{\rm tot}}}
\def\sub{{{\rm sub}}}
\def\obs{{{\rm obs}}}
\def\dep{{{\rm dep}}}

\shorttitle{Unified Model of GRBs}
\shortauthors{Yamazaki, Ioka \& Nakamura}
\begin{document}

\title{A Unified Model of Short and Long Gamma-Ray Bursts, 
X-Ray Rich Gamma-Ray Bursts, and X-Ray Flashes}
\author{Ryo Yamazaki$^{1,2}$, Kunihito Ioka$^{2}$ 
and Takashi Nakamura$^{1}$}
\affil{
$^{1}$Department of Physics, Kyoto University,
Kyoto 606-8502, Japan 
\\
$^{2}$Department of Earth and Space Science,
Osaka University, Toyonaka 560-0043, Japan
}


\begin{abstract}
Taking into account the recent suggestion that a short gamma-ray burst 
(GRB) looks like the first 1 sec of a long GRB, we propose 
that the jet of a GRB consists of multiple sub-jets or 
sub-shells (i.e., an inhomogeneous jet model).
The multiplicity of the sub-jets along a line of sight $n_s$ is 
an important parameter. 
If $n_s$ is large ($\gg 1$) the event looks like a long GRB, 
while if $n_s$ is small ($\sim 1$) the event looks like a short GRB. 
If our line of sight is off-axis to any sub-jets, 
the event looks like an X-ray flash or an X-ray rich GRB.
The log-normal distribution of durations of short and long GRBs
are also suggested in the same model.
\end{abstract}



\section{Introduction}
For the long gamma-ray bursts (GRBs), 
the cosmological distance, the collimated jet,
the massive star progenitor, and the association with the supernova 
are almost established or strongly suggested
\citep[e.g.,][]{m02,zm03}.
However,  for  short GRBs, little is known
since no afterglow has been observed.
The origin of the X-ray flashes (XRFs) also remains
unclear although many models have been proposed
\citep[see][and references therein]{yin04a}.
The observed event rate of  short GRBs is about a third of the long GRBs 
while the observed event rate of  XRFs is also about a third
\citep{He01a,ki02,lamb2003}.
Although there may be a possible bias effect to these statistics, 
in an astrophysical sense, these numbers are the same or  comparable. 
If these three phenomena arise from essentially different origins, 
the similar number of events is just by chance. 
While if these three phenomena are related like Seyfert~1 and 2 galaxies,
the similar number of  events is natural and 
the ratio of the event rate tells us something about
the geometry of the central engine 
\citep{awaki1991,antonucci1993,urry1995}.
In this Letter, we propose a unified model in which
the central engine of short GRBs, long GRBs and  XRFs is the same  
and the apparent differences come essentially from  different viewing angles. 

\section{Unified Model}
It is suggested that  short GRBs are similar to 
the first 1 sec of  long GRBs \citep{ggc03}.
Although  short GRBs are harder than long GRBs \citep{k93},
this difference is mainly due to the difference 
in the low-energy spectral slope while the peak energy is similar
\citep{ggc03}.
Other properties, such as $\langle V/V_{\rm max}\rangle$, 
the angular distribution,
the energy dependence of duration
and the hard-to-soft spectral evolution
of  short GRBs, are also similar to those of long GRBs
\citep{lamb2002}.
If  short GRBs also obey the peak energy-luminosity relation 
found for the long GRBs \citep{y03},
it is suggested that  short and long GRBs have
a similar redshift distribution\footnote{
Even if the afterglows of the short and long GRBs have a similar 
mechanism, the current limits are still consistent with the lack of 
afterglows for short GRBs \citep{h02,lamb2002,kba03}.}
\citep{ggc03}.

\begin{figure}
\plotone{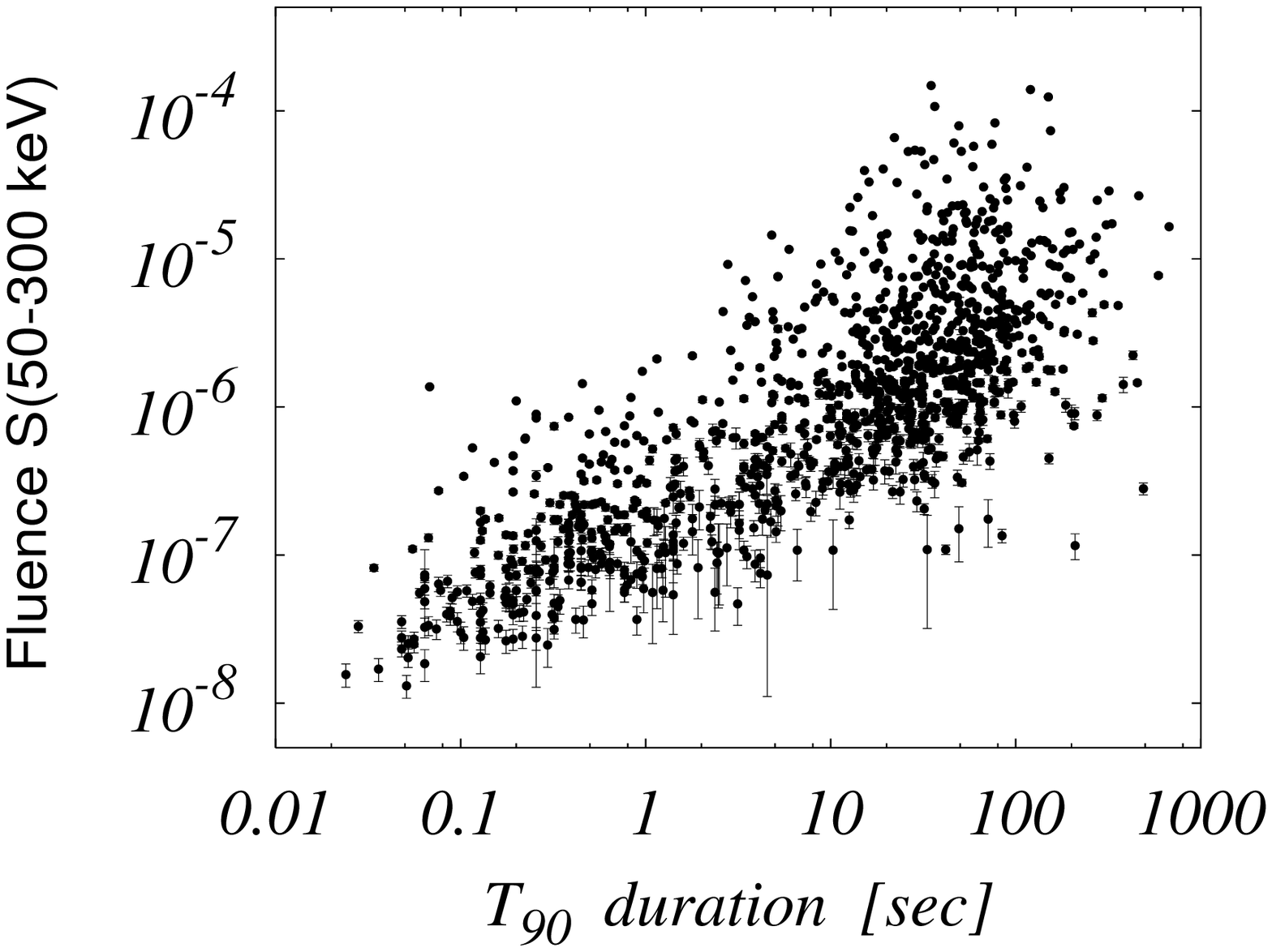}
\caption{
The fluence $S(50-300~{\rm keV})$ as a function of $T_{90}$
duration for BATSE bursts from 4Br catalog.
Courtesy of Drs.~S.~Michikoshi and T.~Suyama.
}
\label{fig_flu_dur}
\end{figure}

These similarities suggest that the difference between  short 
and long GRBs is just the number of pulses, 
and each pulse is essentially the same \citep{rf00}.
As shown in Fig.~\ref{fig_flu_dur},
using 4Br catalogue of BATSE \citep{pac99}, the fluence is roughly 
in proportion to the duration in the range of 0.01 to 1000 sec
\citep[see also][]{balazs2003}.
Thus, we may consider that each pulse is produced
by essentially the same unit or the sub-jet, and the GRB jet consists 
of many sub-jets.
If many sub-jets point to our line of sight, the event looks like 
the long GRB while  if a single sub-jet points to us, 
the event looks like a short GRB.
Since we can observe only the angular size of $\sim \gamma^{-1}$
within the GRB jet with the Lorentz factor $\gamma$,
different observers will see different number of sub-jets
depending on the distribution of sub-jets within the GRB jet.
Since the angular size of a causally connected region is also
$\gamma^{-1}< 0.01 $,
the opening half-angle of a sub-jet can be much smaller than that of
the whole GRB jet ($\sim 0.1$), say $\sim0.02$.

XRFs also appear to be related to GRBs.
Softer and dimmer GRBs smoothly extend to the XRFs
\citep{He01a,ki02,lamb2003,watson04},
while the peak energy-isotropic luminosity/energy relations hold
for GRBs as well as XRFs \citep{s03,y03,a02}.
%
The total energy including the radio afterglow of 
XRF~020903, which has a measured redshift,
might be similar to that of  GRBs \citep{so03}.
Other properties, such as the duration, the temporal structure
and the Band spectrum of the XRFs are also similar to those of the GRBs,
suggesting that XRFs are in fact soft and dim GRBs.
In the sub-jet model, XRFs are naturally expected 
when our line of sight is off-axis to any sub-jets
\citep{nakamura2000,in01,yin02,yin03b,yin04a}.

The origin of sub-jets is not yet clear.
In this Letter, we do not discuss the origin of the sub-jets,
but argue the implications of the sub-jet model.

\section{An Example of Numerical  Simulation of Our Unified Model}
\begin{figure}
\plotone{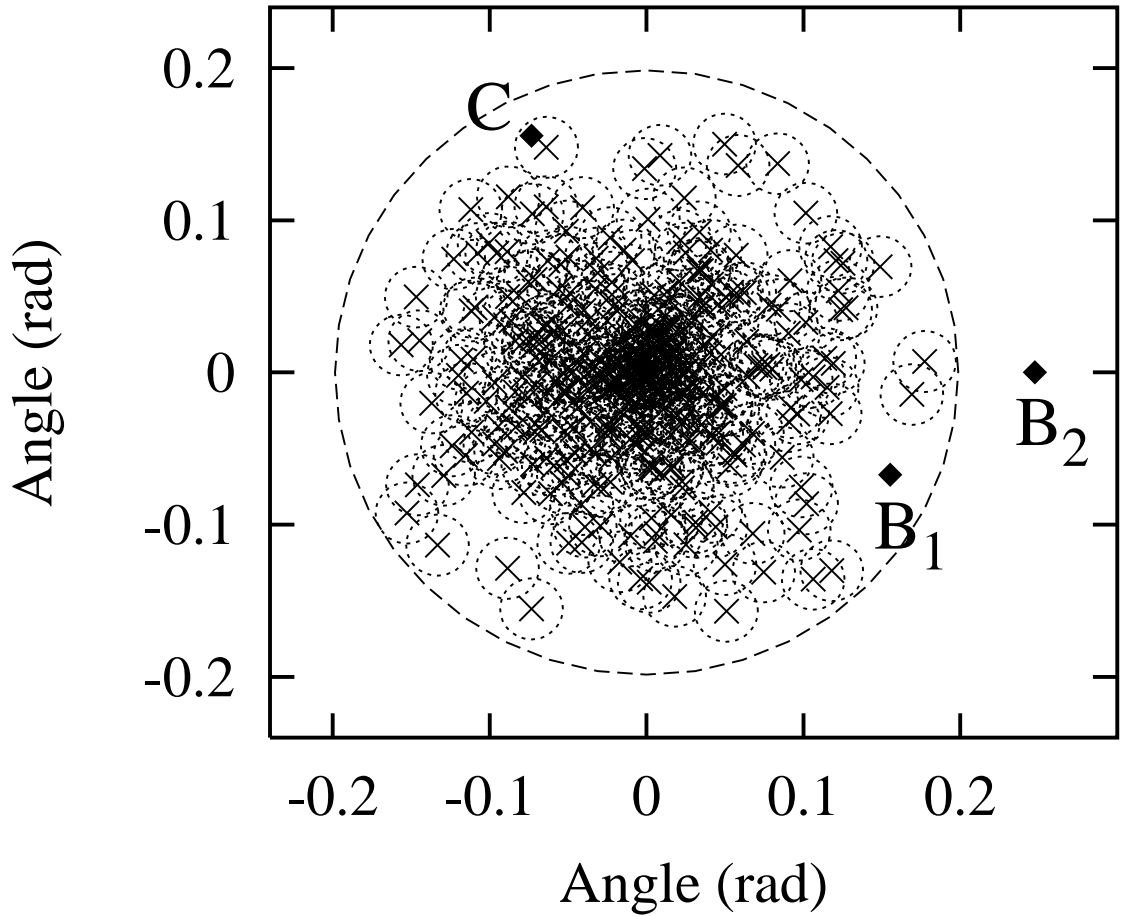}
\caption{
The angular distribution of $N_\tot=350$ sub-jets confined 
in the whole GRB jet in our simulation.
The whole jet has the opening half-angle of 
$\Delta\theta_\tot=0.2$~rad.
The sub-jets have the same intrinsic luminosity,
opening half-angles $\Delta\theta_\sub=0.02$~rad and 
other properties; $\gamma=100$, $r=10^{14}$\,cm, 
$\alpha_B=-1$\,, $\beta_B=-2.5$\,, $h\gamma\nu'=500$\,keV.
The axes and the angular size of sub-jets are represented by crosses
and the dotted circles, respectively. 
 ``A'' represents the center of the whole jet and is hidden by the lines of sub-jets. \
}\label{fig1}
\end{figure}

\begin{figure}
\plotone{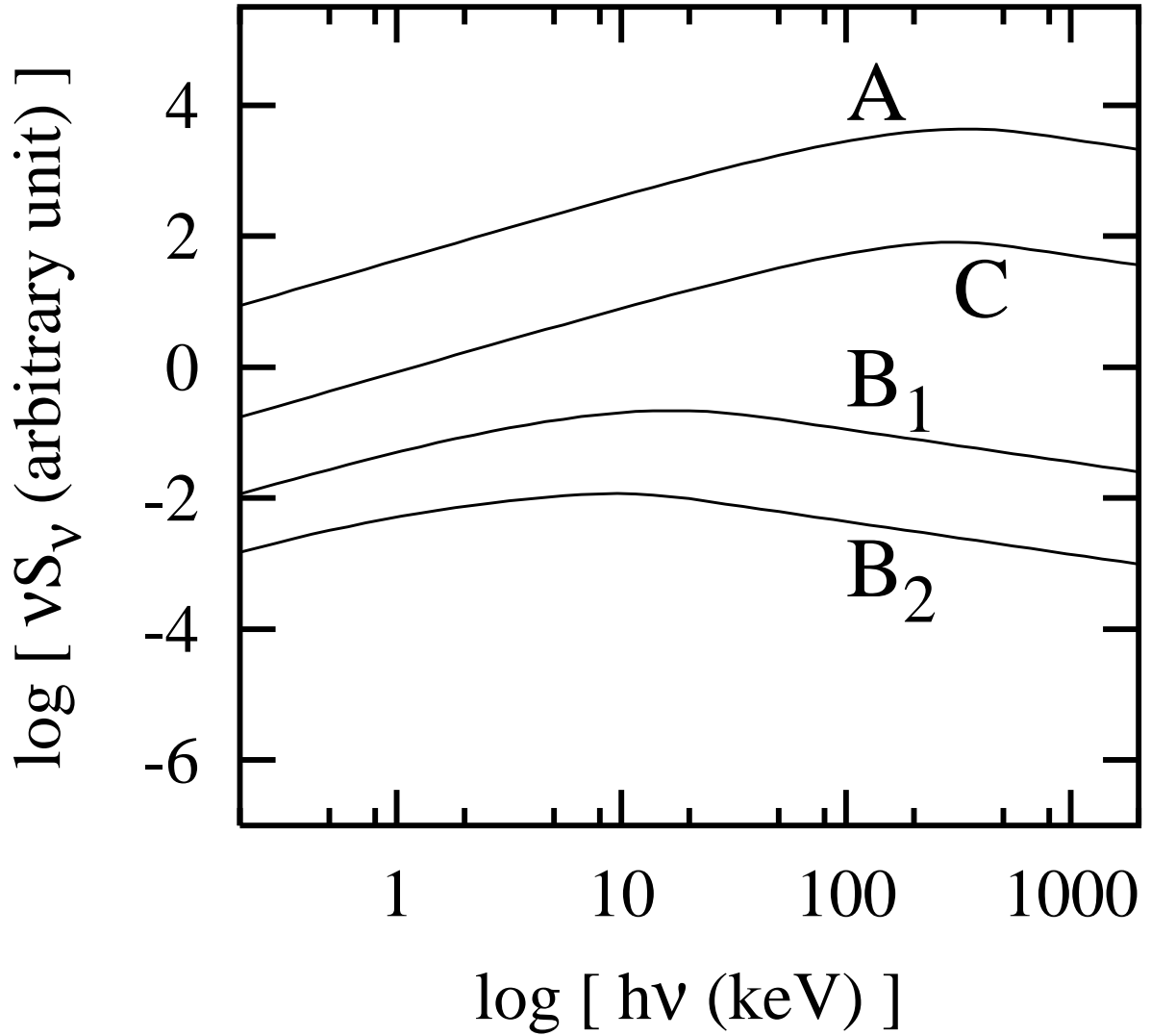}
\caption{
Time-integrated energy spectrum of the emission from the multiple sub-jets
for the observers denoted by 
``A'', ``B$_1$'', ``B$_2$'', and ``C'' in Figure~\ref{fig1}.
The source are located at $z=1$. 
}\label{fig2}
\end{figure}

\begin{figure}
\plotone{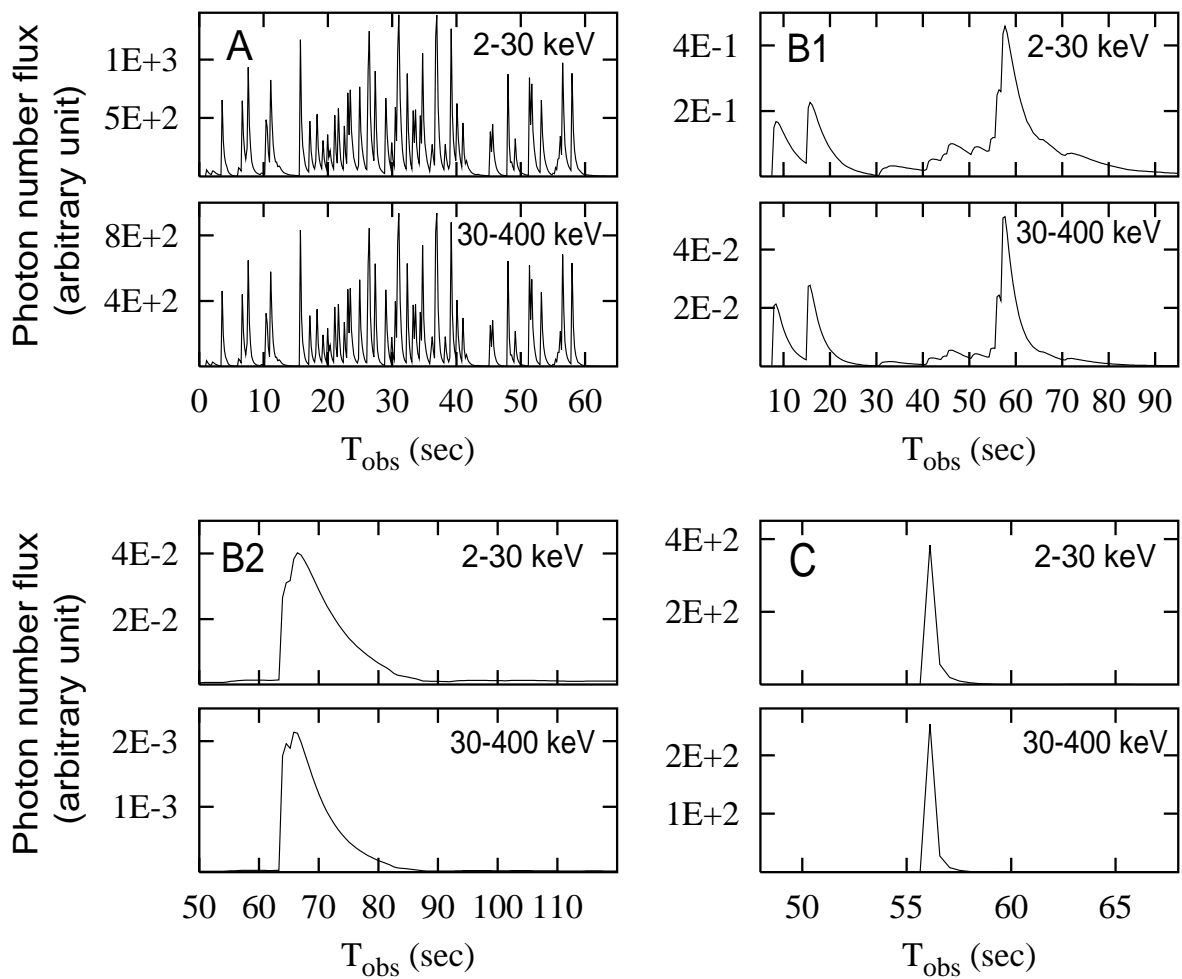}
\caption{
The observed X-ray and $\gamma$-ray light curves from the multiple 
sub-jets,
corresponding the cases ``A''(the upper left),
``B$_1$''(the upper right), ``B$_2$''(the lower left) and
``C''(the lower right) in Figure~\ref{fig1}.
The sources are located at $z=1$. 
}\label{fig3}
\end{figure}

We first show a numerical simulation to demonstrate how
the event looks so different depending on the viewing angle 
in our unified model. 
Let us consider $N_\tot=350$ sub-jets, for simplicity, 
confined in the whole GRB jet
whose axis is the same as a $\vartheta=0$ axis.
For each sub-jet the emission model is the same as in \citet{yin03b}.
Let the opening half-angle of the $j$-th sub-jet 
($j=1,\,\cdots,\,N_\tot$) be $\Delta\theta_\sub^\j$,
while the opening half-angle of the whole jet be
$\Delta\theta_\tot$.
The direction of the observer and the axis of the $j$-th sub-jet 
are specified by $(\vartheta_\obs\,,\varphi_\obs)$
and $(\vartheta^\j,\varphi^\j)$, respectively.
We assume the $j$-th sub-jet departs at time $t_\dep^\j$
from the central engine
and emits at radius $r=r^\j$ and time 
$t=t^\j\equiv t_\dep^\j+r^\j/\beta^\j c$,
where $t$ and $r$ are measured in the central engine frame
and we set $t_\dep^{\scriptscriptstyle (j=1)}=0$.
For simplicity, all sub-jets are assumed to have the same intrinsic
properties, that is 
$\Delta\theta_\sub^\j=0.02$~rad, $\gamma^\j=100$~, 
$r^\j=10^{14}$~cm,
$\alpha_B^\j=-1$, $\beta_B^\j=-2.5$, 
$\gamma h{\nu'}_0^\j=500$~keV and the amplitude 
$A^\j={\rm const}.$ for all $j$.
The departure time of each sub-jet, $t_\dep^\j$ is randomly
distributed between $t=0$ and $t=t_{\rm dur}$,
where $t_{\rm dur}$ is the  active time
of the central engine measured in its own frame and set to
$t_{\rm dur}=30$~sec.
The opening half-angle of the whole jet is set to 
$\Delta\theta_\tot=0.2$~rad as a typical value.
We consider the case in which 
the angular distribution of  sub-jets is given by 
$P(\vartheta^\j,\varphi^\j)\,d\vartheta^\j\, d\varphi^\j\propto
\exp[-(\vartheta^\j/\vartheta_c)^2/2]
\,d\vartheta^\j\, d\varphi^\j$
for $\vartheta^\j<\Delta\theta_\tot-\Delta\theta_\sub$,
where we adopt  $\vartheta_c=0.1$~rad \citep{z03}.
In this case,
sub-jets are concentrated on the $\vartheta=0$ axis (i.e., the
multiplicity in the center $n_s\sim 10$).
For our adopted parameters,  sub-jets are sparsely distributed in the 
range $\vartheta_c\lesssim\vartheta\lesssim\Delta\theta_\tot$,
however,  the whole jet would be entirely filled
if the sub-jets were uniformly distributed 
(i.e., the mean multiplicity $n_s\sim 3$).
Therefore, isolated sub-jets exist near the edge of the whole jet 
with the multiplicity $n_s\ll 1$
and there exists a viewing angle where no sub-jets are launched.
Figures~\ref{fig1}, \ref{fig2} and \ref{fig3} show 
the angular distributions of sub-jets and
the directions of four selected lines of sight, the observed time-integrated
spectra, and the observed light curves in the X-ray and $\gamma$-ray
bands, respectively. Note here in Figure~\ref{fig1}, ``A'' represents 
the center of the whole jet and is hidden by the lines of sub-jets. 

{\it Long GRB}: 
When we observe the source from the $\vartheta=0$ axis
(case~``A''), we see spiky temporal structures
(Fig. 3-A)  and  $E_p \sim 300$~keV
which are typical for the long GRBs.
We may identify  case ``A'' as  long GRBs. 

{\it XRF and X-ray rich GRB}: 
When the line of sight is away from any sub-jets
(cases~``B$_1$'' and ``B$_2$''),
soft and dim prompt emission, i.e.  XRFs
or X-ray rich GRBs are observed with $E_p= 10\sim 20$~keV and $\sim 4$
orders of magnitude smaller fluence than that of   case ``A''
(Fig. 2). 
The burst duration is comparable to that in  case  ``A''. 
These are quite similar to the characteristics of XRFs. 
We may identify the
cases ~``B$_1$'' and ``B$_2$'' as  XRFs or X-ray rich GRBs.

{\it Short GRB}: 
If the line of sight is  inside an isolated sub-jet (case~``C''),
its observed pulse duration is $\sim 50$ times smaller than 
 case ``A'' (Fig.~3-C). 
Contributions to the observed light curve from the other sub-jets
are negligible so that the fluence is  about a hundredth
 of the case ``A''. 
These are quite similar to the characteristics of  short GRBs.
However the hardness ratio 
($=S(100-300~{\rm keV})/S(50-100~{\rm keV})$)  is about 3 
which is smaller than the mean hardness of short GRBs ($\sim 6$). 
\citet{ggc03} suggested that the hardness of  short GRBs
is due to the large low-energy photon index $\alpha_B\sim -0.58$ 
so that if the central engine
launches $\alpha_B\sim -0.58$ sub-jets to the periphery of the core 
where
 $n_s$ is small, we may identify the case ``C'' as the short-hard GRBs.
In other words, the hardness of 3 comes from $\alpha_B =-1$ in our simulation
so that if $\alpha_B\sim -0.58$, the hardness will be 6 or so. 
We suggest here that not only the isotropic energy but also the photon
index may depend on $\vartheta$. Another possibility is that
if short GRBs are the first 1~sec of the activity of the 
central engine, the spectrum in the early time might be
$\alpha_B\sim -0.58$ for both  the sub-jets in the core and the 
envelope. 
This is consistent with a high KS test probability for $E_p$  and 
$\alpha_B$ \citep{ggc03}. These possibilities may
have something to do with the origin of $\alpha_B\sim -1$ 
for the long GRBs. 

{\it X-ray pre/post-cursor}: 
It is quite interesting that in Figure~\ref{fig3},
we see the X-ray precursor at $T_{\rm obs}\sim60$~sec in  ``B$_2$'' 
 and the postcursor at $T_{\rm obs}\sim65$--75~sec in ``B$_1$''.
These can be understood by the model proposed by \citet{nakamura2000}.

\section{Log-normal Distributions}
The total duration of the long and short GRBs are consistent with 
the log-normal distributions \citep{m94}.
In our sub-jet model, these distributions may be naturally expected
as a result of the central limit theorem.
Suppose a certain quantity $q$ is expressed by a product of
random variables
$q=x_{1} x_{2} \cdots x_{n}.$
Then,
$\log q=\log x_{1}+\log x_{2}+ \cdots + \log x_{n}.$
When the individual distributions of $\log x_{i}$ satisfy 
certain weak conditions, the distribution of $\log q$
obeys the normal distribution in the limit of $n\to \infty$ by
the central limit theorem.
However in some cases the log-normal distributions can be achieved
only by a few variables \citep{in02}.
Thus we might say, ``Astrophysically, not $n \to \infty$ but $n=3$
gives the log-normal distribution in practice!''.
This argument may apply to the log-normal distributions
of the peak energy, the pulse fluence and the pulse duration
of  GRBs \citep{in02}.

In our sub-jet model, the short GRBs are due to a single sub-jet.
The pulse duration of a single sub-jet is mainly determined by
the angular spreading time scale and is given by the product
of four variables in the internal shock model \citep{in02} as
$\Delta T_{\rm short}\sim(1+z)(L/c)(\gamma_s/\gamma_m)^2$
where $L$, $\gamma_s$, $\gamma_m$ are the separation of two shells, 
the Lorentz factor of the slow and merged shell in the internal
shock model, respectively.
Therefore the log-normal distribution of the duration of the short GRBs 
may be a natural result of the central limit theorem.

In our unified model, the duration of long GRBs 
is determined by
the interval between pulses $\Delta t=L/c$
times the multiplicity of the sub-jets $n_{s}$.
For a GRB at redshift $z$, the observed duration is given by
the product of three random variables,
$\Delta T_{\rm long}\sim(1+z)(L/c)n_{s}$.
Therefore the log-normal distribution of the  duration
of long GRBs may be realized.  The ratio of the duration of
 long GRBs to  short GRBs is given by
$n_s(\gamma_m/\gamma_s)^2\sim 10^2$. 
Since in the internal shock model
the relative Lorentz factor is not large, 
this equation suggests that $n_s=10\sim 30$ 
which is compatible with the observed number of spikes of the long GRBs.

\section{Discussions}
Let $\Delta\theta_{\rm sub}$, $\vartheta_c$ and $\bar{n}_s$ be the
typical opening half-angle of the sub-jet, the core size of the whole jet
and the mean multiplicity in the core. Then the total number of the
sub-jets ($N_\tot$) is estimated as
$N_\tot=\bar{n}_s(\vartheta_c/\Delta\theta_{\rm sub})^2\sim 10^3$ so
that the total energy of each sub-jet is $\sim 10^{48}$ erg. 
In our model, 
the event rate of long GRBs is in proportion to $\vartheta_c^2$. 
Let $M$ be the number of sub-jets in the 
envelope of the core with a small multiplicity $n_s\ll 1$. 
Then the event rate of short GRBs is in proportion to
$M\Delta\theta_{\rm sub}^2$ so that $M\sim 10$ is enough to explain 
the event rate of short GRBs. 

Of course, the above numerical values are typical ones
and should have a dispersion \citep{ldz03}.
Our core-envelope sub-jet model can have a similar structure to
the two component jet model 
\citep{b03,h03,zwh03,rcr02}
by varying such as  $\bar{n}_s$ and $M$.
However the distribution of sub-jets could also have other 
possibilities, e.g., a hollow-cone distribution like a pulsar, 
a power law distribution, a Gaussian distribution 
\citep{zm02,rossi2002,z03} and so on.

Some observers could see a cold spot with small $n_s$ in the core
to have a small geometrically corrected
energy even if the total energy of the GRBs is the same.
Thus our model may be compatible with the recent claim
that the total kinetic energy has smaller dispersion
than the geometrically corrected $\gamma$-ray energy 
\citep{b03,bfk03}.
The X-ray pre-/post-cursor is also expected
if off-axis sub-jets are ejected earlier (for precursor)
or later (for postcursor) than the main sub-jets \citep{nakamura2000}.
The viewing angle of the sub-jets may also cause
the luminosity-lag/variability/width relations of the GRBs 
including GRB 980425
\citep{yyn03,in01}.
This multiple sub-jet model is an extreme case of the 
inhomogeneous or patchy shell model \citep{kp00,nakamura2000}.  The afterglow variabilities, such as in GRB 021004, 
may arise from the angular energy fluctuations within the GRB jet
\citep{np03,png03}, which might correspond to the
inhmogeneous $n_s$.

Since the core may be regarded as a uniform jet,
our model for the XRFs is analogous to the off-axis uniform jet model
\citep{yin02,yin03b,yin04a}.
However the afterglow could have a different behavior between
the core-envelope sub-jet model and the uniform jet model.
In the uniform jet model, the afterglows of XRFs should resemble
the orphan afterglows that initially have a rising light curve
\citep[e.g.,][]{yin03a,g02}.
An orphan afterglow may be actually observed in XRF 030723 \citep{h03},
but the light curve may peak too early \citep{z03}.
The optical afterglow of XRF 020903 is not observed initially ($<0.9$ days)
but may not be consistent with the orphan afterglow \citep{so03}.
These problems could be overcome by introducing a Gaussian tail
with a high Lorentz factor around the uniform jet \citep{z03}
since the energy redistribution effects may bring the rising light curve
to earlier times \citep{z03,kg03}.
The afterglow of a short GRB is difficult to predict since it
could resemble both the orphan and normal afterglow depending on 
the sub-jet configuration within the envelope.

Since all bursts have the same progenitor,
our model suggests that the short GRBs and the XRFs are also associated 
with  supernovae. 
The radio calorimetry will also give a similar energy to 
 long GRBs because of the same reason.
Our unified model will be refuted if the locations of short GRBs are 
mainly in the halo of the galaxy, as in the coalescing binary neutron star 
model \citep{bkd02}. 

Interestingly our model also predicts off-axis short GRBs or short XRFs.
However these bursts will be difficult to detect since
 short XRFs, which have a multiplicity of $n_{s}\sim 1$, 
will be $\sim 30$ times dimmer than XRFs with $n_{s}\sim 30$.
Note that the short XRFs will be longer than the short GRBs
since the pulse duration grows as the viewing angle increases
\citep{in01,yin02}.
The event rate of  short XRFs will depend on the configuration of the
sub-jets in the envelope.
Further observations are necessary to determine the envelope structure.

\acknowledgements

%
We would like to thank G.~R.~Ricker, T.~Murakami, 
N.~Kawai, A.~Yoshida, and K.~Touma for useful comments and discussions.
This work was supported in part by
a Grant-in-Aid for for the 21st Century COE
``Center for Diversity and Universality in Physics''
and also supported by Grant-in-Aid for Scientific Research
of the Japanese Ministry of Education, Culture, Sports, Science
and Technology, No.05008 (RY),
No.660 (KI),
No.14047212 (TN), and No.14204024 (TN).


\end{document}